# High-Voltage Terminal Test of Test Stand for 1-MV Electrostatic Accelerator

**Sae-Hoon Park[1,2], Yu-Seok Kim[2]**

[1]*KOMAC, Korea Multipurpose Accelerator Complex, Gyenogju 780-904*

[2]*Department of Energy & Environment Engineering, Dongguk University, Gyeongju 780-714*

The Korea Multipurpose Accelerator Complex (KOMAC) has been developing a 300-kV test stand for a 1-MV electrostatic accelerator ion source. The ion source and accelerating tube will be installed in a high-pressure vessel. The ion source in the high-pressure vessel is required to have a high reliability. The test stand has been proposed and developed to confirm the stable operating conditions of the ion source. The ion source will be tested at the test stand to verify the long-time operating conditions. The test stand comprises a 300-kV high-voltage terminal, a battery for the ion-source power, a 60-Hz inverter, 200-MHz RF power, a 5-kV extraction power supply, a 300-kV accelerating tube, and a vacuum system. The results of the 300-kV high-voltage terminal tests are presented in this paper.

Email: psh@kaeri.re.kr

Fax: 054-705-3826

## I. INTRODUCTION

The Korea Multipurpose Accelerator Complex (KOMAC) can provide an ion beam whose energy will be expanded to the megaelectronvolt range for developing an electrostatic accelerator, which is configured to provide a maximum accelerating voltage of 1 MV, a maximum beam current of 10 mA, and various gaseous ions such as nitrogen and oxygen. The most important components for an electrostatic accelerator are a high voltage power supply, an ion source, and an accelerating tube.

The ion source and accelerating tube will be installed in a high-pressure vessel filled with $SF_6$. The ion source in the high-pressure vessel must have a high reliability. It will be tested at the test stand to verify the long-time operating conditions. The test stand has been proposed and developed in order to confirm the stable operating conditions of the ion source at atmospheric pressure. The test-stand voltage is 300 kV, with an operating time of 8 h. For the ion-source power supply, the chargeable battery will be supplied power and charge at night. The test stand comprises a 300-kV high-voltage terminal, a battery for the ion-source power, a 60-Hz inverter, 200-MHz RF power, a 5-kV extraction power supply, a 300-kV accelerating tube, and a vacuum system. The beam-measurement device and beam dump will be installed in the latter part of the accelerating tube [1]. A block diagram of the test stand of the 300-kV high-voltage terminal components is shown in Fig. 1.

The electric field was calculated using POISSON code in order to confirm its strength for a 300-kV high-voltage terminal of the test stand. The test results for the power supply of the RF ion source and the 200-MHz RF power for the ion source are presented.

## II. POISSON CODE FOR 300-kV HIGH-VOLTAGE TERMINAL

The high-voltage terminal comprises 300-kV insulators and a terminal part, and it is supported by the high-voltage insulators. The terminal consists of four insulators with a diameter of 15 cm and height of 83 cm. The high voltage terminal is 86 x 86 x 148 cm in size. After it was assembled, a high-voltage test was conducted. We experimented with a non-equipped corona ring, and discharge occurred at the insulator surface (245 kV).

Corona rings were used to improve the performance of insulator strings. They reduce corona discharges as well as the associated audible noise level. Corona rings also improve the reduction of the electric-field level. They can be used to adjust the voltage distribution along the insulator near the ends of the insulator, thereby reducing the maximum electric field [2].

The high-voltage terminal insulators were simulated by POISSON code in Cartesian coordinates and cylindrical coordinates in order to calculate the electric field at the corona-ring surface and insulator surface and simulate various conditions.

All simulations were conducted using POISSON code. This program solves the electrostatic and electromagnetic problems separately. The program generates a triangular mesh fitted to the boundaries of different materials in the problem geometry. Plotting programs and other postprocessor codes can be used to obtain results in various forms [3].

We preformed the POISSON code simulation and illustrated the potential contours around the insulator and corona ring. Terminals with and without an equipped corona ring were simulated separately to identify the effects of the ring at the insulator surface. We conducted simulations while changing the diameter of the corona ring.

The simulation results of the high-voltage terminal show a simplified half structure with a 300-kV insulator and terminal box (conductor, not equipped with a corona ring) in Cartesian coordinates. The peak value of the electric-field strength was near the junction of the insulator and terminal. For the high-voltage terminal equipped with a corona ring, the peak value of the electric field was at the corona-ring surface (Cartesian coordinates and cylindrical coordinates). Figure 2 illustrates the potential contour of the high-voltage terminal simulation results in Cartesian coordinates. Figure 3 illustrates the electric field along the insulator surface with a corona ring.

We conducted the simulation while changing the diameter of the corona ring in both coordinates systems. The peak values of the electric field at the corona-ring surface are presented in Table. 1. Numerically, the electric field on the corona ring decreases from 57.764 to 38.323 kV/cm as the diameter of the corona ring increases.

The use of a corona ring in an insulator string significantly decreases the electric-field strength on the insulator surface. That is, the potential distribution is uniform when a corona ring is used. As the corona tube diameter increases, the electric-field strengths of the corona-ring surfaces decrease. The peak value of the electric-field strength under the corona-ring surface decreases as the corona tube diameter increases. We conducted a high-voltage test with a corona ring ( 25.4), and discharge occurred with an audible noise under the corona-ring surface at the peak value point (300 kV). We manufactured a 50.8-mm-diameter corona ring and assembled it at the test stand, as shown in Fig. 4.

### III. RF ION SOURCE

An RF ion source has been operated at a different site in KOMAC. Another RF ion source will be installed at the test stand for the 1-MV electrostatic accelerator. A test stand of the RF ion source consists of an RF system, an impedance-matching network, an RF ion source, a vacuum system, and a beam diagnostic system. Its specifications are a 200-MHz frequency, a 5-kV extraction voltage, and a vacuum system. The RF system includes a power supply, voltage-controlled oscillator (VCO), solid-state amplifier (SSA) operating at 200 MHz, circulator, directional coupler, and dummy load. A switched-mode power supply (SMPS, 600 W) provides electric power to the VCO and SSA. The VCO is controlled by varying the resistance and 200 MHz RF transferred to the SSA. Block diagrams of the RF system are shown in Fig. 5. The RF power is supplied by the SSA, whose maximum output power is 200 W. In order to confirm the operating conditions of the VCO and SSA, we measured the RF power using a power meter (Agilent E4416A) and the analyzed VCO by a spectrum analyzer (Agilent

E4403B). The results are shown in Fig. 6. The VCO was operated at 200 MHz with a controlling adjustable resistance (13.73 V). A circulator transfers the reflected RF power to a dummy load, which absorbs the reflected RF power.

The chargeable battery (12 V, 400 Ah) was installed at the second stage of the high-voltage terminal. The battery was connected to a 60-Hz inverter (12 V) that supplied electric power to the RF system chassis. The RF system chassis comprises the SMPS, VCO, SSA and attenuator. The SMPS supplied power to the VCO and SSA with a given supply voltage. The attenuator controlled the RF system power. We tested the battery life using a signal generator and the RF system chassis. The battery-performance test results are shown in Tables 2 and 3.

A 200-MHz RF ion source was installed at the test stand. It consisted of an air variable capacitor comprising a loading and tuning capacitor, a 1-turn coil, a permanent magnet, a shielding box, and an electrode [4]. The plasma was confined by an axial magnetic field produced by permanent magnets placed around a 20-mm pyrex tube. The magnetic field of the permanent magnet was 0.1 T at the center. Impedance matching was adjusted using L-network air variable capacitors. The extraction-hole diameter was 4 mm, and the distance between the electrodes was 7 mm; this was modified for the beam extraction. We used hydrogen and pressure maintained at 1.0 E-5 torr to generate the plasma. The hydrogen plasma is shown in Fig. 7. The extraction power supply was controlled by a signal generator using an optical convertor at a high-voltage rack. The extracted beam current obtained at a 5-kV extraction voltage was 0.29 mA. The high-voltage power supply (Cockcroft-Walton) for the high-voltage terminal was operated from 0 to 300 kV. A beam extraction experiment for the test stand was performed, and the beam current was measured by a Faraday cup in the chamber. The results are shown in Table 4.

## IV. CONCLUSION

In conclusion, the KOMAC has been developing a test stand for an ion source for the 1-MV electrostatic accelerator. Most importantly, the 300-kV high-voltage terminal for POISSON simulation results is introduced. The electric-field strengths at the corona-ring surfaces decrease as the corona tube diameter is increased. A high-voltage experiment with a 50.8-mm-diameter corona ring for a 300-kV test stand was completed. Moreover, the battery for the ion source power was tested at 200 MHz with the signal generator and a VCO. The battery was used to supply power to the RF system over 8 h. Hydrogen plasma was generated by the 200-MHz RF system. Beam extraction was performed for the test stand with a beam current of 300 μA.

## ACKNOWLEDGEMENT

This work was supported by the KOMAC (Korea Multipurpose Accelerator Complex) operation fund of KAERI by MSIP (Ministry of Science, ICT and Future Planning).

## REREFENCES

[1] Yong-Sub Cho, Kye-Ryung Kim, Chan-Young Lee, Basic Design Study on 1-MV Electrostatic Accelerator for ion irradiation, Transactions of the Korean Nuclear Society Spring Meeting, Jeju, Korea, May 29-30, 2014.

[2] W. Sima, F.P.Espino-Cortes, Edward A. Cherney, and Shesha H. Jayaram, IEEE, September 2004 , pp. 480-483.

[3] Los Alamos Accelerator Code Group, User's Guide for the Poisson/Superfish group of Codes, Los Alamos National Laboratory, LA-UR-87-126, (1987).

[4] Jeong-tae Kim, Kwang-mook Park, Dong-Hyuk Seo, Han-Sung Kim, Hyeok-Jung Kwon, Yong-Sub Cho, Hydrogen Plasma Generation with 200-MHz RF Ion Source, Transactions of the Korean

Nuclear Society Autumn Meeting, Pyeongchang, Korea, October 30-31, 2014.

**TABLE CAPTIONS**

Table 1. Peak value of electric field in Cartesian coordinates and cylindrical coordinates, with changing diameter of corona ring

Table 2. Chargeable battery test results for RF system with signal generator

Table 3. Chargeable battery test results for RF system with VCO

Table 4. 300-kV test stand beam-extraction test results

**TABLES**

Table 1

| Diameter of corona ring (mm) | Electric field in Cartesian coordinate (kV/cm) | Electric field in cylindrical coordinate (kV/cm) |
|---|---|---|
| 25 | 25.697 | 57.784 |
| 38 | 21.523 | 44.453 |
| 50 | 19.032 | 38.323 |

Table 2

| Used quantity of battery (%) | Operating time (h) | Signal generator 200 MHz (dBm) | Power meter (W) | Power meter (dBm) |
|---|---|---|---|---|
| 26.7 | 8.5 | -4.65 | 82.2 | 49.15 |
| 26.6 | 7.5 | -4.15 | 100 | 50.01 |
| 24.5 | 9.0 | -4.05 | 101 | 50.05 |
| 22.3 | 6.4 | -4.05 | 101 | 4.05 |

Table 3

| Used quantity of battery (%) | Operating time (h) | Power meter (W) | Power meter (dBm) |
|---|---|---|---|
| 33.4 | 9.3 | 197 | 52.93 |
| 33.4 | 9.5 | 198 | 52.96 |
| 11.2 | 3.8 | 195 | 52.90 |

Table 4

| High voltage (kV) | Leak current (mA) | Extraction power supply voltage (kV) | Extraction power supply current (mA) | Beam current (μA) |
|---|---|---|---|---|
| 130 | 0.1 | 5.0 | 0.29 | 250 |
| 140 | 0.2 | 5.0 | 0.29 | 280 |
| 145 | 0.2 | 5.0 | 0.29 | 300 |

**FIGURE CAPTIONS**

Fig. 1. Block diagram of test stand for 300-kV high-voltage terminal

Fig. 2. POISSON simulation results for high-voltage terminal in Cartesian coordinates; without equipped corona ring (left) and with equipped corona ring (right)

Fig. 3. Calculated electric field along 300-kV insulator surface with corona ring

Fig. 4. Test stand: chamber (left), 300-kV high-voltage terminal (middle), and Cockcroft-Walton high-voltage power supply (right)

Fig. 5. Block diagram of RF system for RF ion source

Fig. 6. VCO analyzed by spectrum analyzer at 200 MHz

Fig. 7. Hydrogen plasma at test stand

**FIGURES**

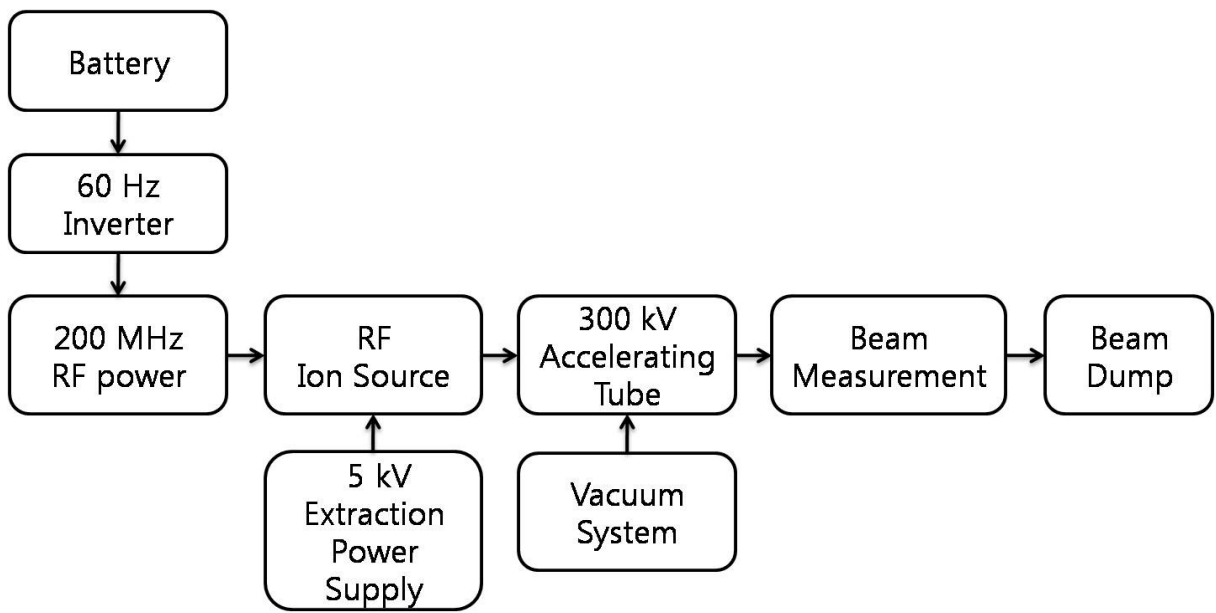

Fig. 1

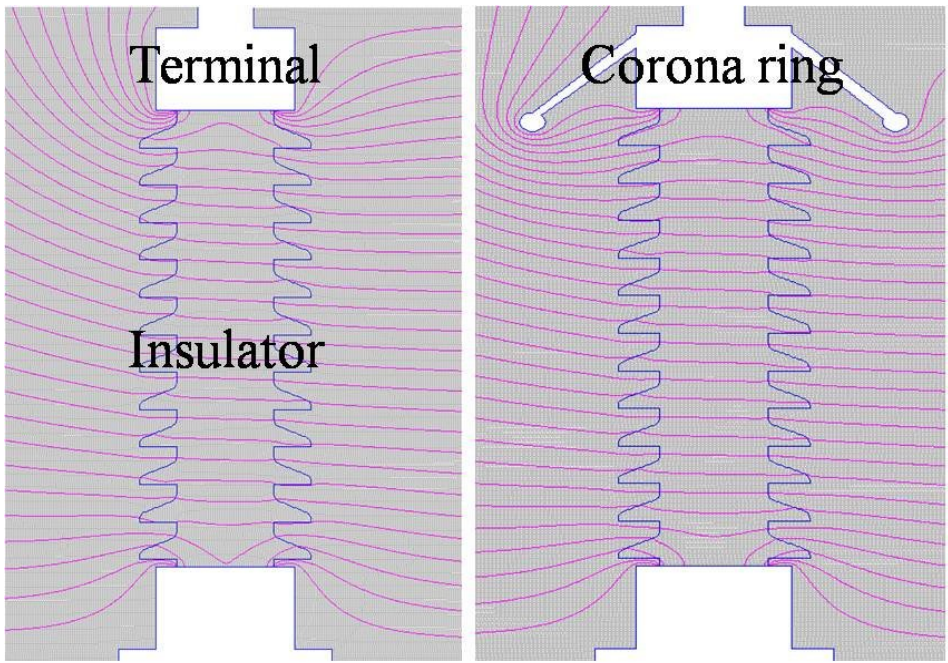

Fig. 2

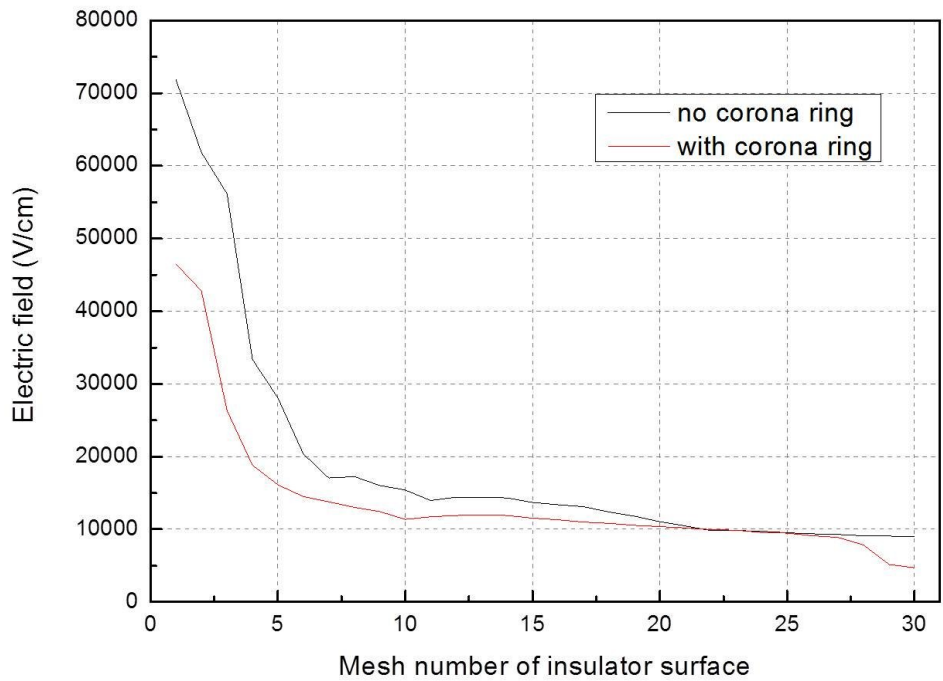

Fig. 3

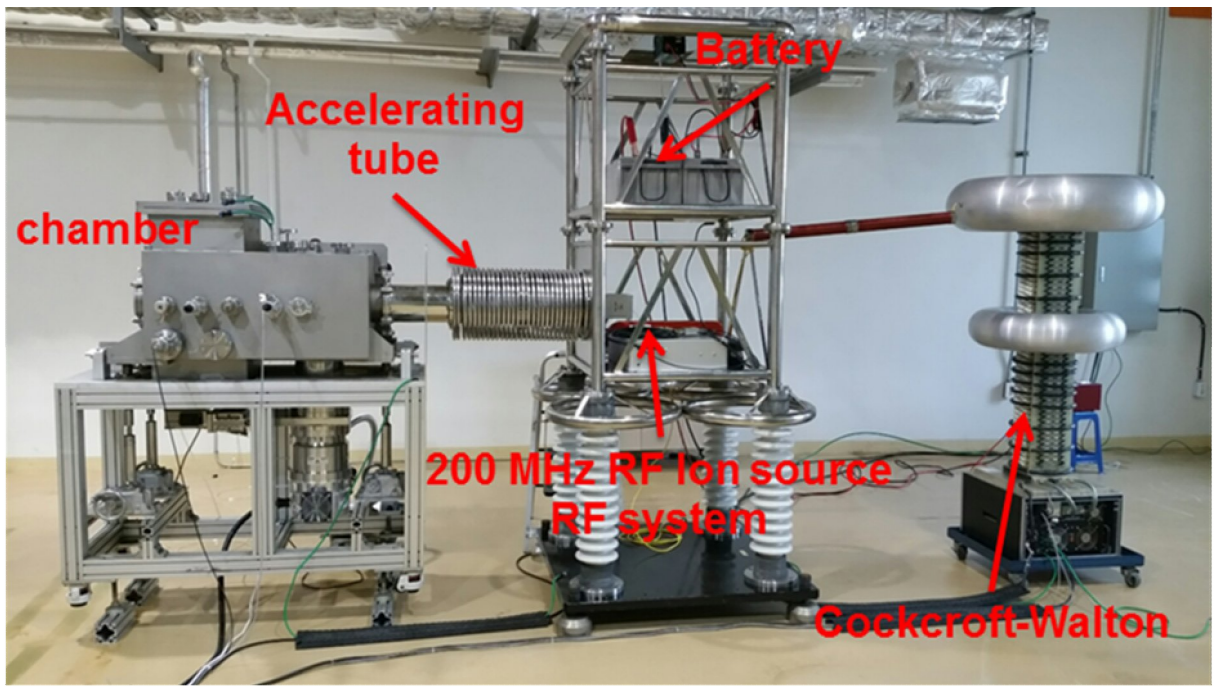

Fig. 4

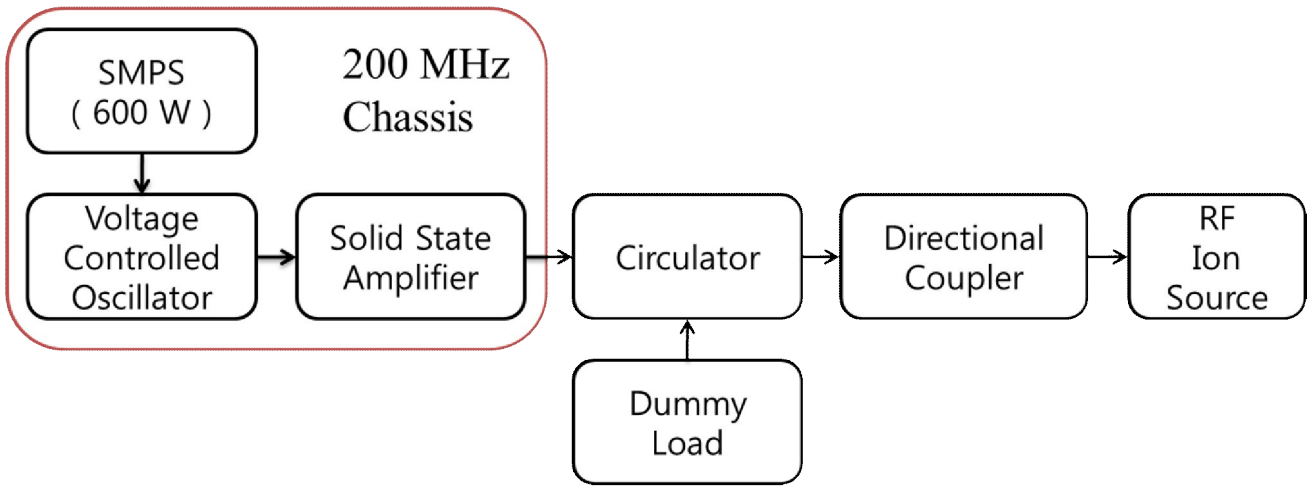

Fig. 5

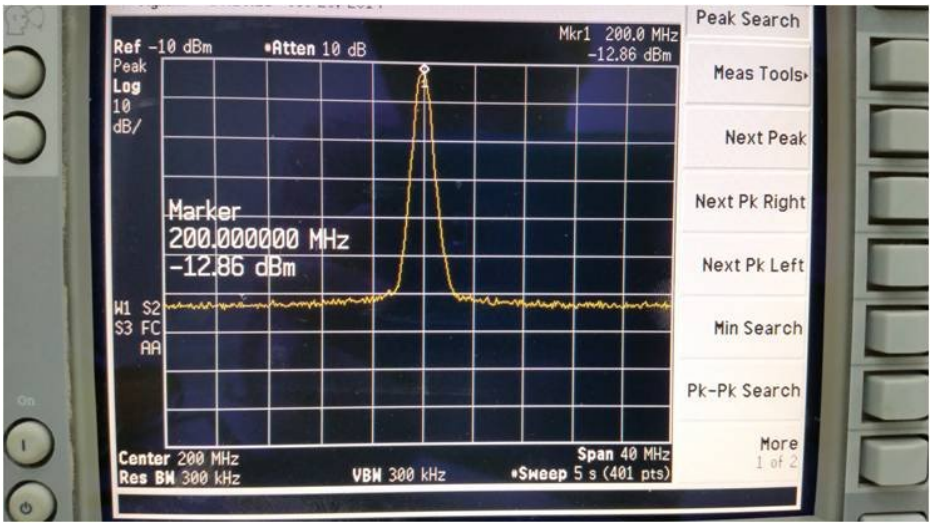

Fig. 6

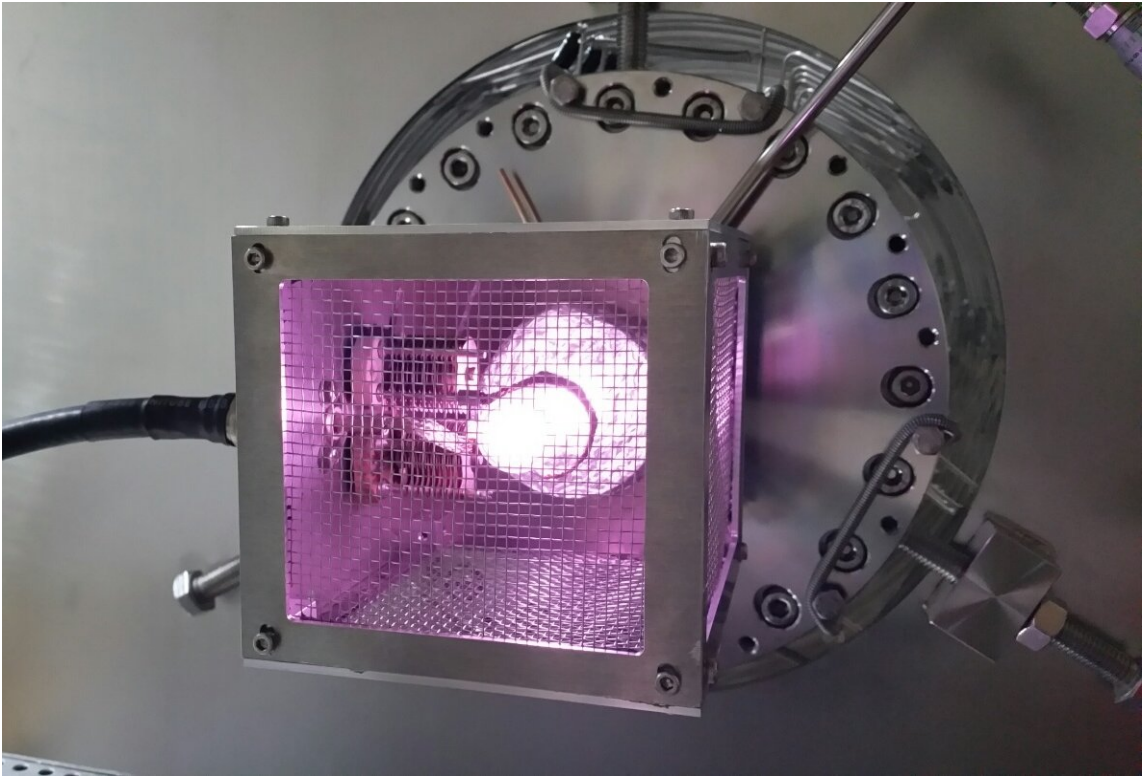

Fig. 7

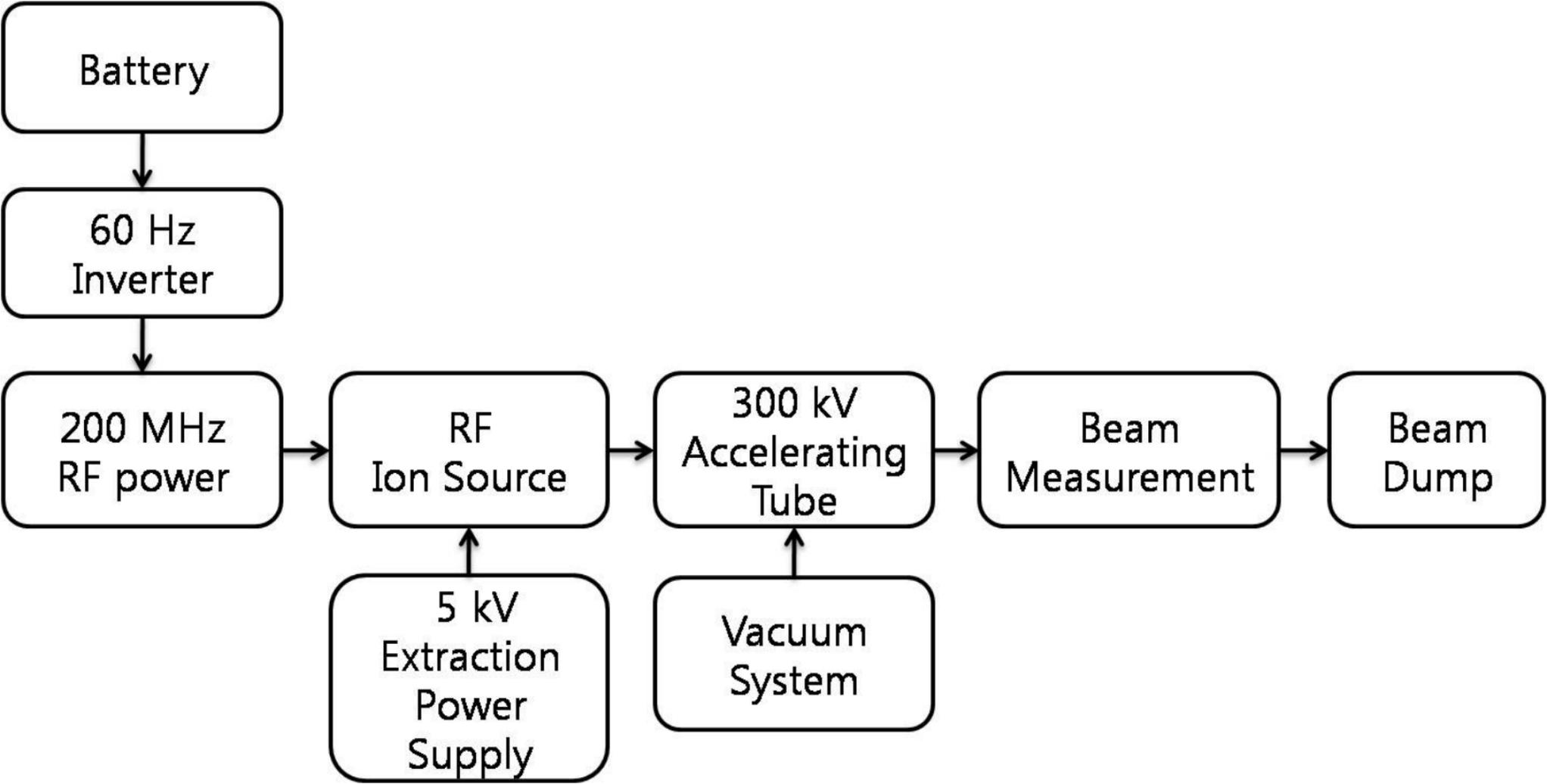

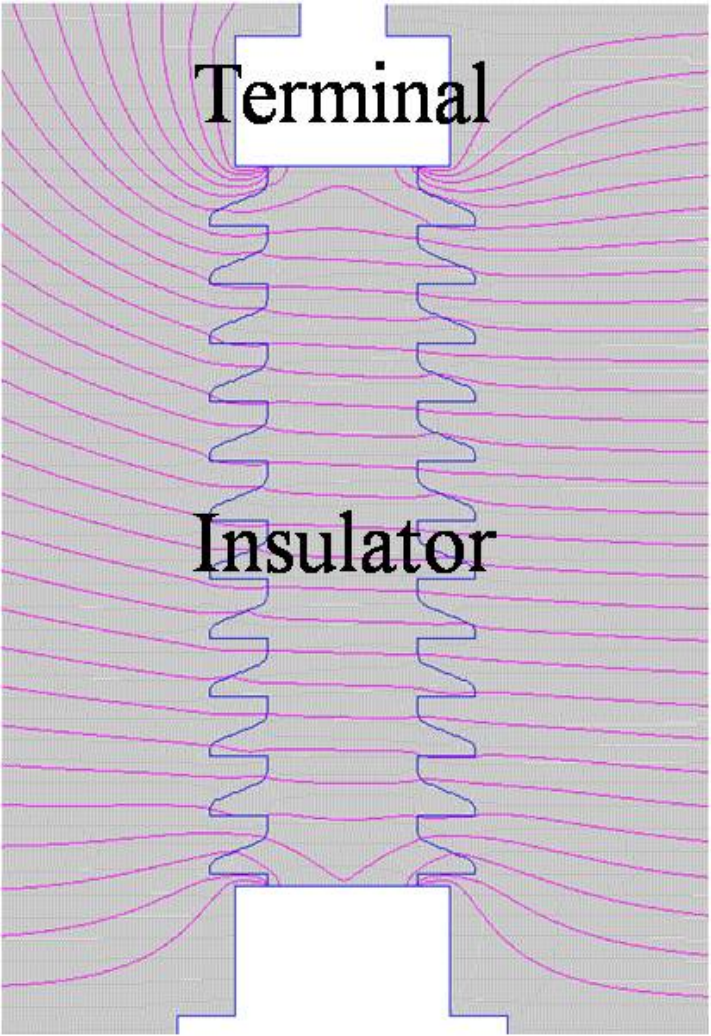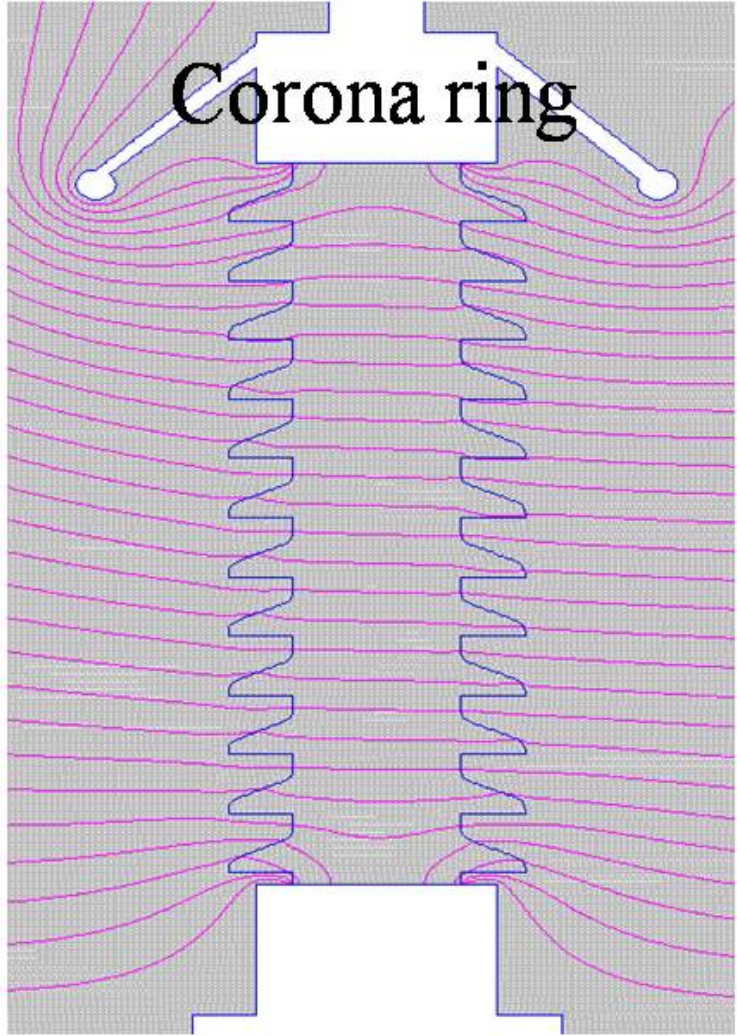

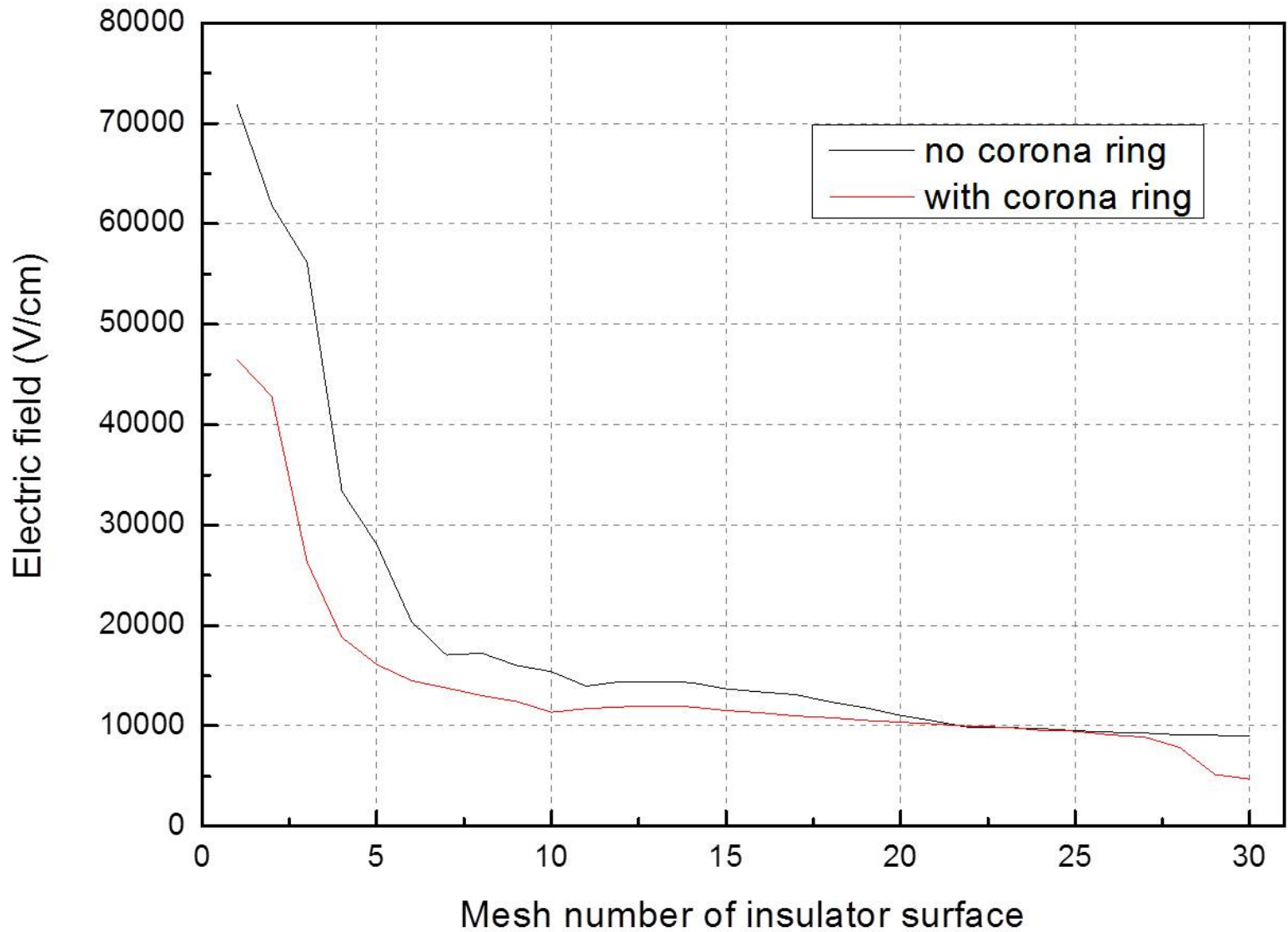

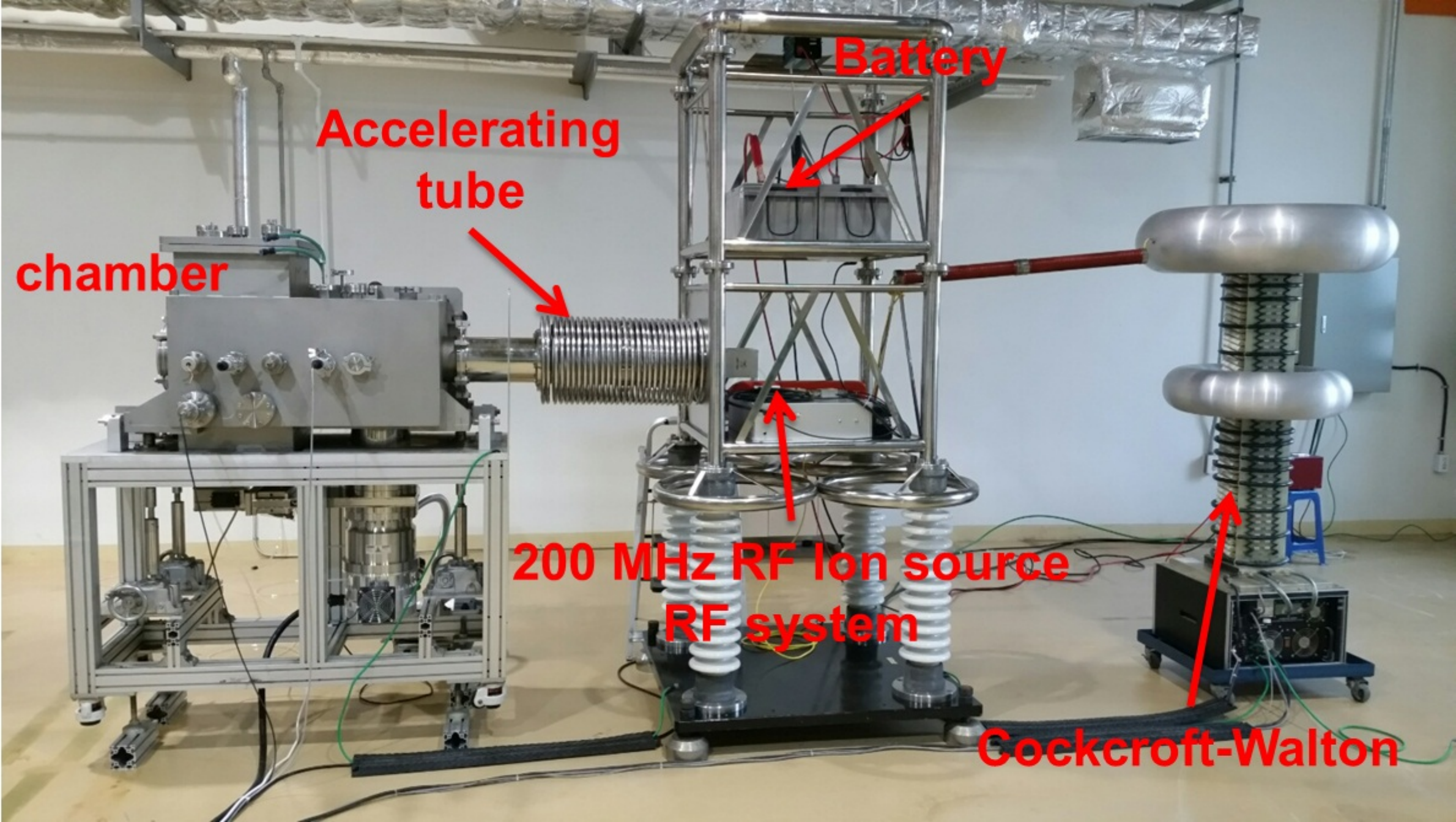

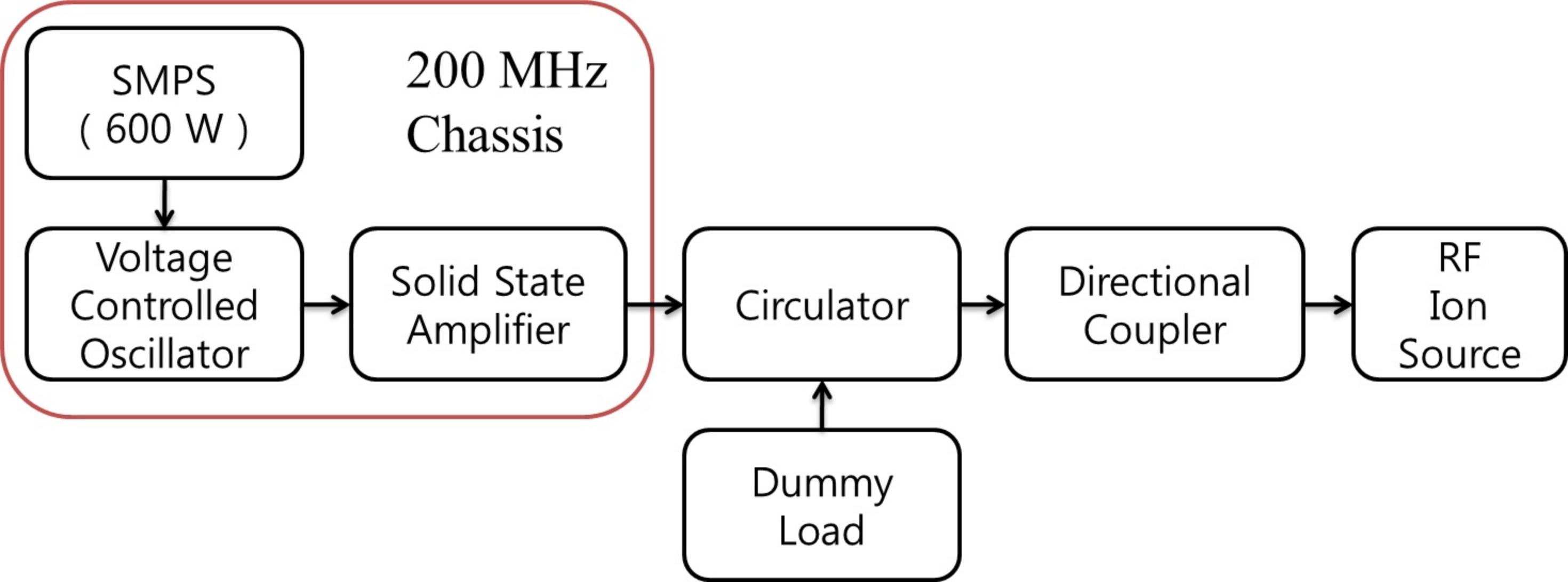

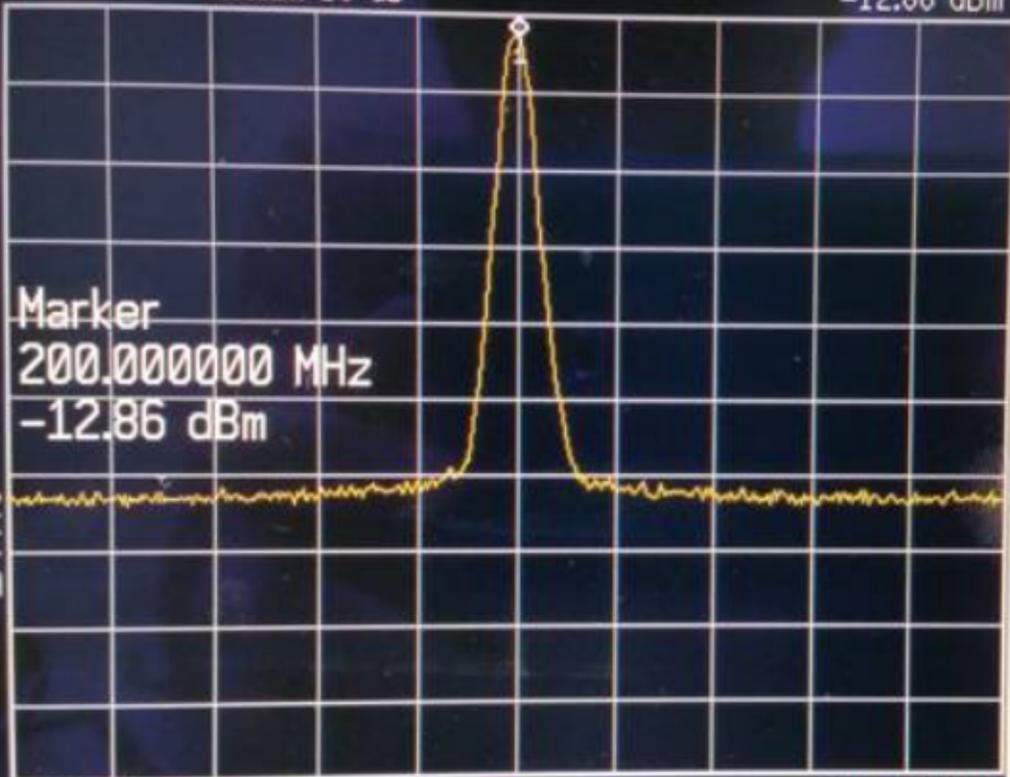

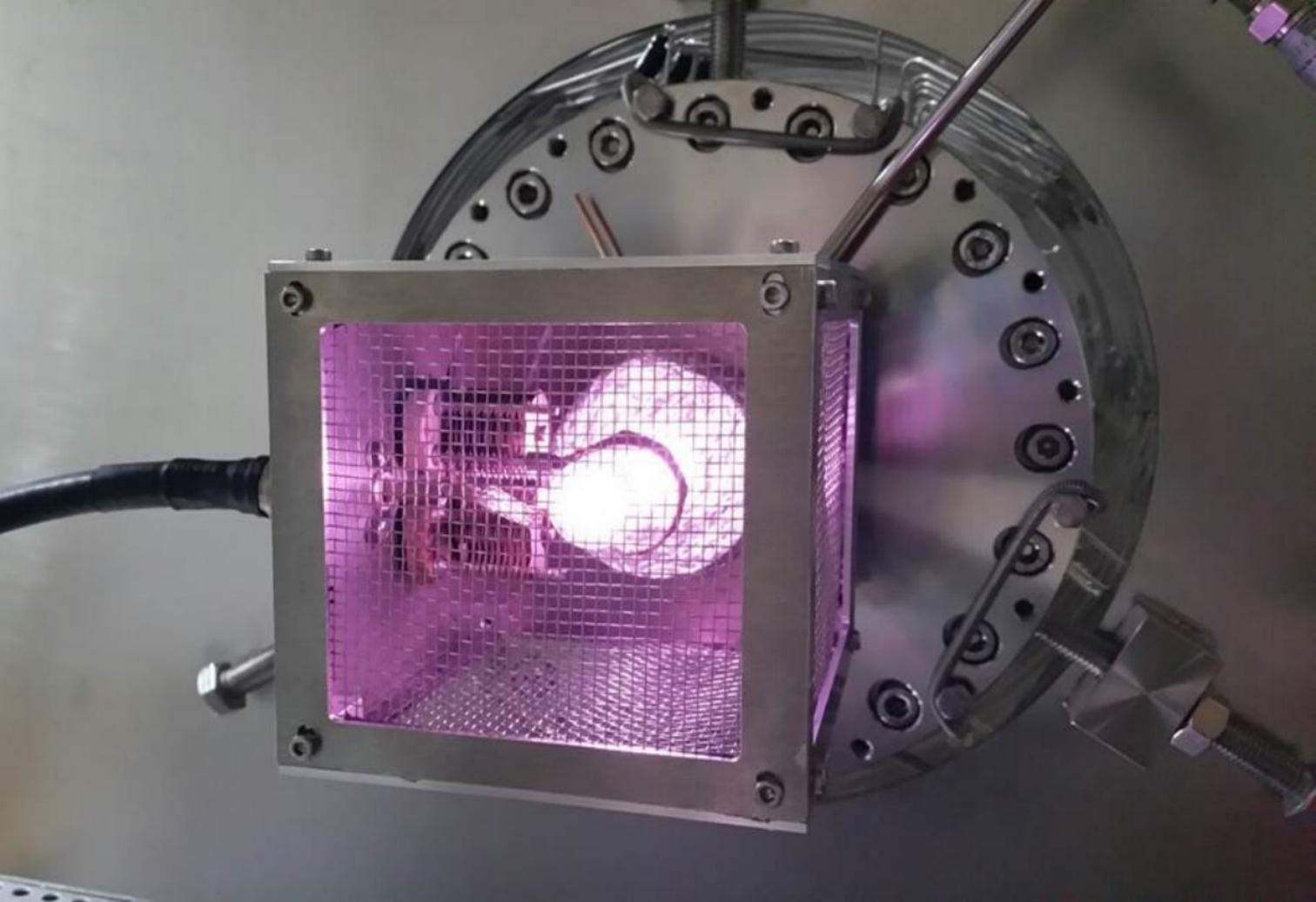